# Green-light p-n Junction Particle Inhomogeneous Phase Enhancement of MgB$_2$ Smart Meta-Superconductor

Yao Qi, Duo Chen, Yongbo Li, Chao Sun, Qingyu Hai, Miao Shi, Honggang Chen and Xiaopeng Zhao *

Smart Materials Laboratory, Department of Applied Physics,

Northwestern Polytechnical University, Xi'an 710129, China;

* Correspondence: xpzhao@nwpu.edu.cn

**Abstract:** Improving the critical temperature ($T_C$), critical magnetic field ($H_C$), and critical current ($J_C$) of superconducting materials has always been one of the most significant challenges in the field of superconductivity, but progress has been slow over the years. Based on the concept of injecting energy to enhance electron pairing states, in this study, we have employed a solid-state sintering method to fabricate a series of smart meta-superconductors (SMSCs) consisting of p-n junction nanostructures with a wavelength of 550 nm, doped within an MgB$_2$ matrix. Experimental results demonstrate that compared to pure MgB$_2$ samples, the critical transition temperature ($T_C$) has increased by 1.2 K, the critical current ($J_C$) has increased by 52.8%, and the Meissner effect ($H_C$) shows significant improvement in its diamagnetic properties. This phenomenon of enhanced superconducting performance can be explained by the coupling between superconducting electrons and evanescent waves.

**Keywords:** smart meta-superconductor; MgB$_2$; green light p-n junction inhomogeneous phase; electroluminescent; injecting energy; electron-evanescent waves coupling; smart superconductivity

## 1. Introduction

Since the discovery of superconductivity, there has been a persistent effort to find practical superconducting materials with high critical temperatures ($T_C$) and high current-carrying capabilities. However, progress in research has been slow over the years. In recent years, novel superconducting systems and phenomena have gained attention, such as copper-oxide superconductors [1], pnictides superconductors [2], iron-based superconductors [3,4], and the graphene magic-angle superconductivity phenomenon [5,6]. Particularly, copper oxide and iron-based superconductors not only exhibit higher critical transition temperatures but also possess superconducting pairing mechanisms that differ from conventional low-temperature superconductors. In Cu-based and Fe-based high-temperature superconductors, the intrinsic superconducting layer/charge reservoir layer interfaces share similarities. In Cu-based oxide high-temperature superconducting materials, the CuO$_2$ superconducting layers are interleaved with oxide charge reservoir layers such as BaO, SrO, LaO, and SmO [4,7,8]. The charge reservoir layers primarily consist of atom substitutions that generate charge carriers, which are transferred to the superconducting layers at the interfaces. This maintains the purity of the superconducting layers, thereby avoiding/reducing impurity scattering and facilitating the realization of macroscopic superconductivity. A similar phenomenon of charge transfer occurs at the FeSe/TiO$_2$ interface, where TiO$_2$ acts as the charge reservoir layer. The bending of the interface energy bands induces charge transfer, leading to the widely recognized interface enhancement effect [9]. Understanding the superconducting mechanisms of these systems holds great significance for the development of practical superconducting materials.

However, with the discovery of superconductivity in H$_3$S at 203 K temperature [10], the focus has once again shifted back to conventional superconductors. In 2019, the Drozdov research group reported that LaH$_{10}$ exhibited a remarkably high critical transition temperature ($T_C$) of up to 250 K at a pressure of 150 GPa [11], breaking the $T_C$ limit for superconducting materials and bringing superconductivity into the realm of room temperature. The characteristics of these superconductors can be described using the Bardeen-Cooper-Schrieffer and Migdal-Eliashberg theories. In principle, materials rich in hydrogen and carbon can provide the high frequencies required in the phonon spectrum, as well as strong electron-phonon interactions [12,13], resulting in higher critical transition temperatures. However, the practical advancement of such materials is severely limited by the extreme high-pressure conditions required to maintain their stable structures.

In addition, manipulating the properties of conventional superconducting materials has been a focus of research efforts. Researchers at the Max Planck Institute in Germany have used ultrafast laser pulses to excite materials, thereby inducing superconductivity in a transient timeframe typically on the order of picoseconds or nanoseconds [14], opening up a new pathway for studying superconducting mechanisms. Another important approach that has emerged in recent years is the manipulation of material properties through the artificial design of material structures, allowing for the realization of unique or even non-existent "anomalous" material properties in nature [15-17]. By controlling the characteristics of heterostructures, we can achieve tunable quantum phenomena, with the most striking examples being the fascinating physical effects observed in twisted bilayer graphene [5,6,18]. Furthermore, heterostructures based on oxide materials [19-22] and two-dimensional van der Waals materials have also served as prominent platforms for demonstrating tunable quantum phenomena [23]. These platforms provide opportunities for tailoring the properties of superconducting materials and exploring novel quantum states, leading to advancements in the field of superconductivity.

Superconducting materials are widely used as building blocks in material structures, not only because of their intrinsic properties but also due to their proximity effect, which relates to the transfer of Cooper pairs from a superconductor to another material at the interface [24,25]. Inspired by these ideas, our group has developed a novel class of smart meta-superconductors (SMSCs) composed of traditional superconducting materials as the matrix and electroluminescent (EL) inhomogeneous phase particles as the building units [26-34]. Y$_2$O$_3$:Eu$^{3+}$ and Y$_2$O$_3$:Eu$^{3+}$+Ag EL materials were introduced into the traditional superconductor MgB$_2$, resulting in the formation of smart meta-superconductors.

When measuring the critical temperature ($T_C$) of SMSCs using a four-probe method, an external electric field can stimulate EL in the inhomogeneous phase, thereby enhancing the Cooper pairing and leading to macroscopic changes in $T_C$. Additionally, SMSCs possess the unique feature of being able to adjust and improve $T_C$ through external electric field stimulation, which is not achievable with traditional second-phase doping [31-33]. We believe this is because the superconducting particles, acting as microelectrodes, excite EL in the inhomogeneous phase under the influence of the external electric field, and the energy injection promotes the formation of electron pairs.

The research results demonstrate that the smart meta-superconductors constructed with rare-earth oxide electroluminescent materials as building units effectively enhance the $T_C$, $J_C$, and $H_C$ of traditional superconducting materials. However, limitations such as low electroluminescent intensity, short luminescence lifetime, and the requirement for high external electric fields hinder the improvement of their superconducting performance. Recently, we have achieved an increase in $\Delta T$ up to 0.8 K and a transition temperature improvement to 39.0 K by incorporating AlGaInP red p-n junction particles inhomogeneous phase, and we propose that the coupling between evanescent waves generated by electroluminescence and superconducting electrons enhances the critical transition temperature [35].

$MgB_2$, as a simple binary compound, was discovered to exhibit superconductivity with a critical temperature of 39 K in 2001 [36]. It has since been a highly active area of research. In comparison to commercially used low-temperature superconductors like NbTi ($T_C$ = 9.7 K) and $Nb_3Sn$ ($T_C$ = 18.3 K), $MgB_2$ offers a higher $T_C$, allowing it to operate within the temperature range of liquid hydrogen and significantly reducing operational costs. Furthermore, compared to high-$T_C$ cuprate superconductors such as $YBa_2Cu_3O_{7-\delta}$, $Bi_2Sr_2Ca_2Cu_3O_{10+\delta}$, and $HgBa_2Ca_2Cu_2O_{8+\delta}$, $MgB_2$ exhibits the ability to carry high current across grain boundaries, eliminating the weak-link problem often observed in high-temperature superconductors. Additionally, $MgB_2$ possesses a large coherence length, making it more prone to the introduction of flux pinning centers, thereby improving its current-carrying performance in high magnetic fields. Therefore, research on $MgB_2$ holds significant value for the development of high-performance and practical superconducting materials.

The present study introduces p-n junction electroluminescent inhomogeneous phase particles with a central wavelength of 550 nm to achieve energy injection and improve electron pairing. Bulk polycrystalline samples were synthesized using a solid-state sintering method, consisting of $MgB_2$ + $x$ wt% p-n junction nanostructures electroluminescent inhomogeneous phase (x = 0.3-1.2). The inhomogeneous phase had particle sizes ($\varphi$) of 0.5 μm, 1.0 μm, 1.5 μm, 2 μm, 3 μm, 4 μm, 5 μm, and 6 μm. The materials were subjected to comprehensive physical property measurements and analysis to carefully investigate the influence of the particle size and doping content of the p-n junction electroluminescent inhomogeneous phase on the structure and superconducting performance of the constructed smart meta-superconductors. The results indicate that, compared to pure $MgB_2$ samples, there is minimal change in the lattice structure of the materials. The critical transition temperature ($T_C$) has increased by 1.2 K, the critical current ($J_C$) has increased by 52.8%, and the diamagnetic properties of the Meissner effect ($H_C$) have significantly improved.

## 2. Model

Figure 1 depicts the model of MgB$_2$ SMSCs constructed using polycrystalline MgB$_2$ as the raw material. The gray polyhedra represent the polycrystalline MgB$_2$ particles, while the green particles represent p-n junction nanostructures with a central wavelength of 550 nm. These p-n junction nanostructures are dispersed as inhomogeneous phase within the MgB$_2$ particles. The introduction of the inhomogeneous phase inevitably lowers the critical transition temperature ($T_C$) of MgB$_2$, primarily due to the non-superconducting nature of the doped inhomogeneous phase, similar to the MgO impurity phase in MgB$_2$. For convenience, the decrease in $T_C$ after the introduction of the inhomogeneous phase is referred to as the impurity effect [26-28]. However, the addition of luminescent inhomogeneous phase has been proven to be an effective method for improving the $T_C$ of MgB$_2$. For example, the introduction of Y$_2$O$_3$:Eu$^{3+}$ and Y$_2$O$_3$:Eu$^{3+}$+Ag can induce electroluminescence effects and enhance $T_C$ [29-33]. There is a clear competition between the impurity effect and the EL excitation effect of the inhomogeneous phase. When the electroluminescence effect dominates, $T_C$ is improved ($\Delta T_C > 0$). Otherwise, the introduction of inhomogeneous phase decreases $T_C$ ($\Delta T_C < 0$). Therefore, to obtain samples with high $T_C$, it is essential to minimize the impurity effect and enhance the EL excitation effect. The superconductivity of smart meta-superconductors can be improved and adjusted by adding EL inhomogeneous phase [32,33,35]. It is well known that changes in $T_C$ are often related to changes in electron density [38,39]. However, under the current preparation conditions, the inhomogeneous phase exists only between the MgB$_2$ particles and does not react with MgB$_2$. Furthermore, diffusion between the inhomogeneous phase and MgB$_2$ particles is difficult, and the electron density cannot be significantly altered. Therefore, the electron density is not a key adjustable parameter that affects the variation of $T_C$. During the measurement process, the applied electric field forms a local electric field within the superconductor and stimulates the emission of photons through EL excitation of the inhomogeneous phase, which is beneficial for the enhancement of Cooper pairs and the variation of $T_C$ [35]. However, considering that photons may disrupt Cooper pairs, further investigation is needed to understand the mechanism behind the $T_C$ variation. Subsequently, based on experimental results, we will use EL excitation of the inhomogeneous phase to explain this phenomenon.

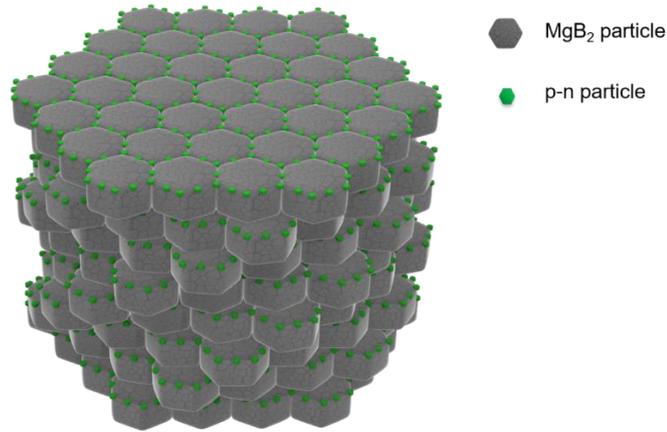

**Figure 1.** MgB$_2$ SMSCs model diagram.

## 3. Experiment

*3.1. Preparation of p-n Junction Luminescent Particles*

We use a commercial green light-emitting diode (LED) epitaxial wafer with a center wavelength of 550 nm, mainly composed of GaN semiconductor structure. This LED structure was manufactured by Xiamen Qianzhao Optoelectronics Co., Ltd. It was grown on a sapphire substrate using metal-organic chemical vapor deposition method, with the activation temperature of the p-type layer exceeding 1300 degrees Celsius. The emissive layer was separated from the substrate and mechanically processed, yielding eight different types of inhomogeneous phase particles with thicknesses of Φ=0.5 μm, 1.0 μm, 1.5 μm, 2 μm, 3 μm, 4 μm, 5 μm, and 6 μm, all measuring 1.7 μm in thickness. These particles consist of a three-layer nanostructure: a p-type semiconductor layer (300 nm thick), an active layer (150 nm thick), and an n-type semiconductor layer (1200 nm thick). We record the untreated LED epitaxial wafer as Sample 1, and the 2 μm p-n junction luminescent particles obtained through the peeling process are labeled as Sample 2. The fabricated 2 μm p-n junction luminescent particles were subjected to high-temperature treatment in a tube furnace under an argon atmosphere, using the same sintering temperature and process as that of MgB$_2$ sintering (850°C for 10 minutes, followed by 650°C for 60 minutes). These treated particles are referred to as Sample 3. Additionally, to ensure the complete activity of the p-n junction luminescent particles in the sintered samples, a control experiment was conducted by subjecting the p-n junction luminescent particles to a heat treatment at 850°C for 120 minutes in an argon atmosphere, referred to as Sample 4. The electro-luminescent performance testing and characterization methods for the four samples are consistent with those used for rare-earth luminescent particles. The emission curves for the various samples under applied voltage <10V and current <10mA are shown in Figure 2.

The rare-earth oxide electro-luminescent particles used in Figure 2(a) were prepared by our research group [33,34]. It can be observed that the emission intensity of the p-n junction luminescent particles is significantly higher than that of the rare-earth oxide electroluminescent particles. During the continuous luminescence performance test conducted for over 2000 hours, no decrease in luminescence intensity was observed. Figure 2(b) illustrates the emission intensity curves of green light p-n junction luminescent particles under different processes. The experimental results indicate that both the luminescent particles treated with the $MgB_2$ sintering process and the p-n junction luminescent particles subjected to more stringent processing retain their complete activity, with no decrease in major performance parameters such as emission wavelength and intensity. The high emission intensity, high-temperature resistance, and long life make the p-n junction luminescent particles an ideal material for the inhomogeneous phase elements in smart meta-superconductor composite material.

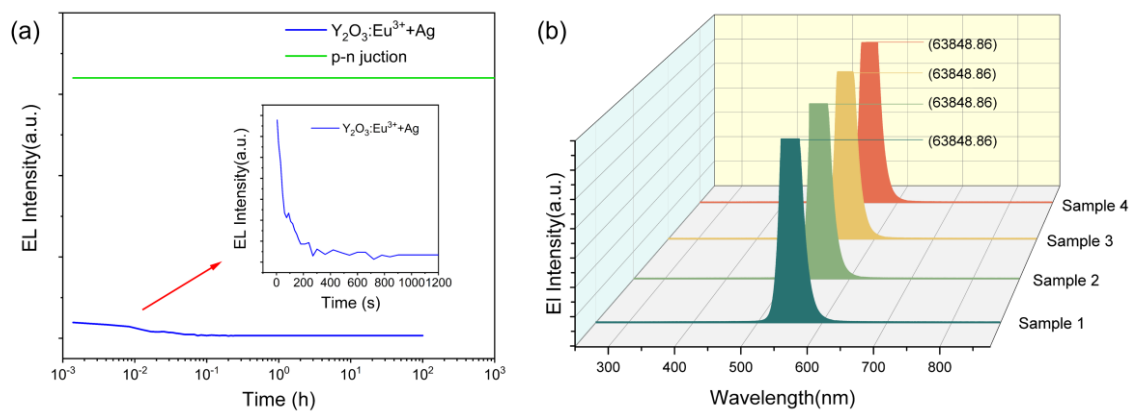

**Figure 2.** Luminescence intensity and lifetime test curves of p-n junction particles and rare earth oxide particles in green light wavelengths

### 3.2. Preparation of MgB₂ Superconductor and Inhomogeneous Phase Samples

The $MgB_2$ used in the experiment was purchased from Alfa Aesar with a purity of 99% and a particle size of 100 mesh (150 micrometers). A certain amount of $MgB_2$ powder was placed in a stainless steel standard sieve with a mesh size of 500 (30 micrometers), and the powder was uniformly dispersed through the sieving process to obtain $MgB_2$ powder with an average particle size below 30 micrometers. 300 mg of the $MgB_2$ powder and the corresponding mass fraction of p-n junction luminescent heterostructure particles were accurately weighed and dispersed in 20ml and 5ml of ethanol solution. The dispersed solutions were subjected to ultrasonic irradiation at 300W power and 40 kHz frequency for 20 minutes, followed by stirring on a magnetic stirrer at a speed of 400 rpm. During the stirring process, the luminescent heterostructure particle solution was gradually added dropwise to the $MgB_2$ particle solution using a 20 μl pipette. After completing the titration, the resulting mixture was stirred for an additional 10 minutes to ensure thorough mixing. Ultrasonic treatment was performed for 20 minutes to prevent particle agglomeration. Finally, the mixture was transferred to a petri dish and dried in a vacuum drying oven at 60°C for 4 hours to obtain a black mixed powder. The obtained black powder was thoroughly ground in an agate mortar and pressed into cylindrical block-shaped samples with a diameter of 11 mm and a thickness of 1.2 mm using a pressure of 14 MPa. The pressed samples were placed in a small tantalum container and then transferred to a vacuum tube furnace. Under a high-purity Ar atmosphere, the samples were slowly heated to 850°C at a heating rate of 5°C/min in the tube furnace, held for 10 minutes, followed by cooling to 650°C at a cooling rate of -5°C/min for 1 hour of sintering, and finally cooled to room temperature to obtain the corresponding samples [25,27,30].

*3.3. Physical phase identification of samples*

The phase identification of the samples was carried out using X-ray diffraction (XRD) with Cu Kα radiation (Germany, Bruker D8 Advance) with a wavelength of 0.15406 nm, an acceleration voltage of 40 kV, and a current of 40 mA. The microstructure of the samples was investigated using a field emission scanning electron microscope (SEM) (United States, FEI Verios G4), and the elemental composition and distribution of the samples were examined using an energy-dispersive X-ray spectrometer (EDS) (United States, ThermoFisher Thermo NS7).

*3.4. Critical Transition Temperature Measurement*

The R-T (resistance-temperature) curve of the samples at low temperatures was measured using the four-probe method, with a distance of 1 millimeter between the four probes. A closed-cycle cryostat, produced by Advanced Research Systems, was used to provide the low-temperature environment (down to a minimum temperature of 10 K). The testing current (1-100 mA) was supplied by a high-temperature superconductor characterization system manufactured by Shanghai Qianfeng Electronic Instrument Co., Ltd., and the voltage measurements were performed using a Keithley nanovoltmeter. The test temperature was controlled using a Lake Shore cryogenic temperature controller. The entire measurement process was conducted under vacuum conditions.

*3.5. Measurement of Critical Current Density and Meissner Effect*

The samples were placed in a low-temperature medium, and the current-voltage (I-V) characteristics were measured using the four-probe method in zero magnetic fields. A certain amount of direct current was passed through two wires connected to the prepared sample, while the other two wires were used to measure the voltage across the sample using a Keithley digital nanovoltmeter. Indium wires were used to connect the sample and the wires, and the distance between the two voltage wires of all samples was 1 millimeter. When the current I passing through the sample exceeded a certain value, the superconducting state was disrupted and transitioned to the normal state. This current is known as the critical transport current of the superconductor. Typically, the critical current density ($J_C$) in a superconducting system is determined by I-V measurements at different temperatures (below the initial transition temperature $T_{C, on}$), with a voltage criterion of 1 μV/cm [34, 42-44]. Throughout the testing process, the shapes, sizes of the samples, and the distances between the current and voltage wires remained unchanged. Subsequently, the prepared samples underwent DC magnetization testing [34, 45]. The samples were slowly cooled in a 1.8 mT magnetic field parallel to the plane, and data were collected during the heating process. All samples exhibited complete diamagnetism.

## 4. Results and Discussion

Figure 3 presents the X-ray diffraction (XRD) pattern of the MgB$_2$ sample doped with 0.9 wt% of p-n nanoscale luminescent inhomogeneous phase particles with a particle size of 2 μm. The black dots represent the experimental data, while the red and blue lines represent the calculated fitted intensity and the difference between the experimental data and the fitted intensity, respectively. The green scale lines indicate the expected Bragg peak positions of the main phase, MgB$_2$ (91.25%). The minor phases, MgO (8.29%) and GaN (0.46%) are represented by the red and purple scale lines, respectively. The color map represents the thermal map representation of the XRD data for the sample. The pattern reveals that the composition of the sintered sample is mainly MgB$_2$, with small amounts of MgO and trace amounts of GaN. No other new phases or components are observed, indicating the stability of the GaN structure during the high-temperature sintering process, without undergoing chemical reactions or elemental substitution with the main phase.

We evaluated the volume percentage of the MgO phase formed in the samples based on the sum of relative X-ray peak intensities and presented it in Table 1. It can be observed that a small amount of MgO phase is inevitably formed in the material, which reduces the critical temperature ($T_C$) of the pure MgB$_2$ sample. However, these small-sized MgO impurities can act as pinning centers, leading to an increase in the critical current density of MgB$_2$ [46]. Additionally, Table 1 provides information on the lattice parameters of all the samples. We observed that the lattice parameters of pure MgB$_2$ were found to be a = 3.08532 Å and c = 3.52317 Å, which closely match the standard MgB$_2$ lattice data. When we conducted doping experiments with inhomogeneous phase particles of varying content and particle size, the lattice parameters of the samples did not exhibit significant shifts or changes, and their XRD diffraction peaks did not show noticeable trends in movement. This indicates that our inhomogeneous phase doping particles are situated between MgB$_2$ particles without causing significant element substitution or perturbations to the MgB$_2$ lattice. The absence of new peaks in the XRD diffraction patterns further supports this observation. From the data in Table 1, we observed that as the doping concentration of the samples increased, the Full Width at Half Maximum (FWHM) systematically increased. This suggests that the crystallinity of the samples decreases due to doping, validating the negative impact of doping on sample performance [47].

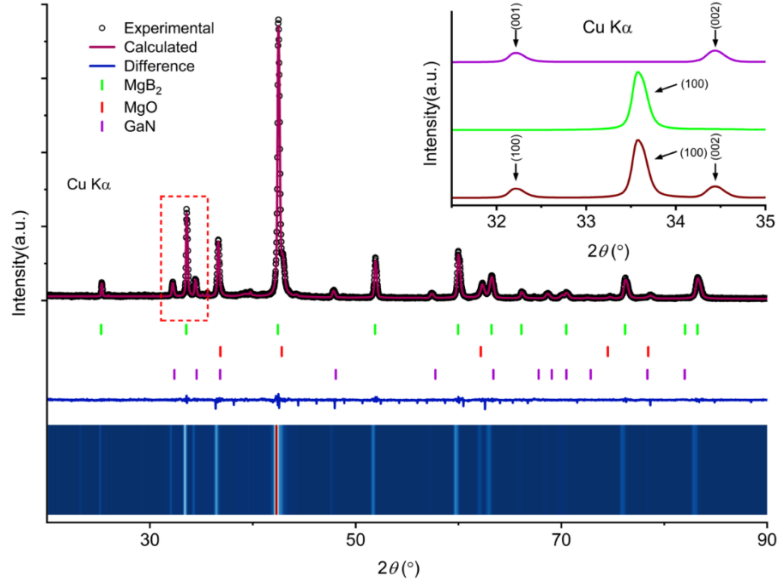

**Figure 3**. XRD studies of the p-n junction nanostructured electroluminescent inhomogeneous phase doped MgB$_2$ samples. (Rietveld refinement of the X-ray powder diffraction data collected at 300 K with Cu Kα radiation.)

**Table 1.** Lattice Parameters, Grain Size, FWHM, MgO Content, and $T_C$ of Samples with Different Doping Concentrations.

| X | GaN size (μm) | a (Å) | c (Å) | Grain size (μm) | FWHM(110) | MgO (%) | $T_C$ (K) |
|---|---|---|---|---|---|---|---|
| 0 | 0 | 3.08532 | 3.52317 | 0.375 | 0.173 | 8.13 | 38.0 |
| 0.4 | 2.0 | 3.08547 | 3.52324 | 0.382 | 0.181 | 8.22 | 37.8 |
| 0.5 | 2.0 | 3.08546 | 3.52327 | 0.377 | 0.184 | 8.21 | 38.2 |
| 0.8 | 1.5 | 3.08523 | 3.52330 | 0.367 | 0.185 | 8.27 | 38.8 |
| 0.8 | 2.0 | 3.08544 | 3.52329 | 0.373 | 0.190 | 8.30 | 38.8 |
| 0.9 | 2.0 | 3.08533 | 3.52334 | 0.381 | 0.193 | 8.29 | 39.2 |
| 0.9 | 3.0 | 3.08528 | 3.52322 | 0.383 | 0.199 | 8.25 | 39.0 |
| 1.0 | 2.0 | 3.08535 | 3.52324 | 0.378 | 0.199 | 8.35 | 38.6 |
| 1.2 | 2.0 | 3.08530 | 3.52319 | 0.382 | 0.207 | 8.24 | 38.2 |

Figure 4 illustrates the microstructure of the doped sample. Figure 4a and 4e show scanning electron microscope (SEM) images of samples doped with 0.9 wt% content of inhomogeneous phase particles with a particle size of 2 μm and pure MgB$_2$, respectively. It can be observed that the particle diameter of the final samples is smaller than 2 μm, and they exhibit a uniformly sized and densely packed structure. The introduction of the inhomogeneous phase does not appear to have a significant impact on the particle size and grain connectivity of the samples. EDS analysis confirms that the luminescent inhomogeneous phase components are uniformly distributed throughout the sample.

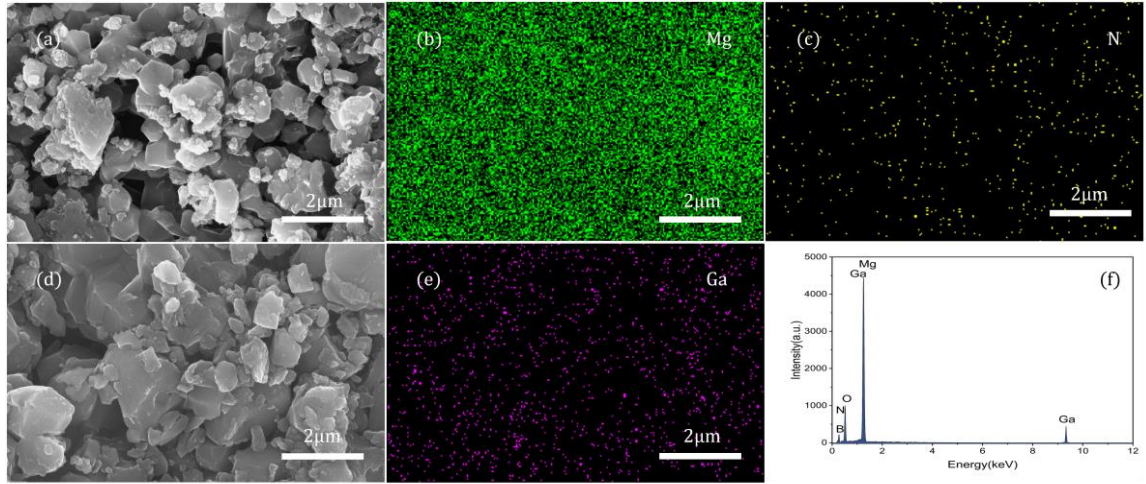

**Figure 4.** (a)(d) SEM diagram of the sample after sintering, (b)(c)(e) EDS mapping of Mg, N, Ga, (f) Elemental surface scan of the sample.

Figure 5 depicts the relationship between the resistivity and temperature ($T_C$) of the MgB$_2$ + $x$ wt% p-n junction electroluminescent inhomogeneous phase particle-doped sample. The doping concentration of inhomogeneous phase components and the particle size are two key factors influencing the performance of SMSCs samples. We prepared eight different specifications of inhomogeneous phase particles with thicknesses of Φ=0.5 μm, 1.0 μm, 1.5 μm, 2 μm, 3 μm, 4 μm, 5 μm, and 6 μm, all with a uniform thickness of 1.7 μm. A series of doped samples with varying doping concentrations were prepared to investigate the optimal doping concentration for each particle size of luminescent particles and the influence of p-n junction nanoscale electroluminescent inhomogeneous phase doping with different particle sizes on the performance of SMSCs samples. The performance characterization results are shown in Figure 5.

In Figure 5a, the resistance-temperature ($R$-$T$) curves of MgB$_2$ samples doped with p-n junction luminescent inhomogeneous phase particles of different diameters ranging from 0.5 μm to 6 μm are presented for the temperature range of 10-300 K. The doping concentrations used for each diameter correspond to the optimal values required to achieve the highest critical transition temperature. The black curve in Figure 5a represents the resistivity variation of the pure MgB$_2$ sample with temperature ($\rho$-$T$), indicating a resistivity $\rho$(300K) of 65 μΩ·cm at room temperature. As the temperature decreases, the resistivity gradually decreases and sharply drops to zero near the superconducting transition temperature, as shown in Figure 5b. Figure 5b is a zoomed-in view of the $R$-$T$ curve from Figure 5a, revealing that the resistivity of the sample starts to rapidly decrease as the temperature reaches 38.0 K, indicating the gradual transition from the normal state to the superconducting state. The superconducting transition temperature ($T_C$) for the pure MgB$_2$ sample is determined to be 37.2-38.0 K with a transition width ($\Delta T_C$) of 0.8 K. The residual resistivity $\rho$(40K) is measured to be 20 μΩ·cm, and the residual resistivity ratio ($RRR=\rho(300K)/\rho(40K)$) is calculated to be 3, showing good agreement with the results reported in the literature [48,49].

From Figure 5a, it can be observed that as the diameter of the inhomogeneous phase particles decreases,

the optimal doping concentration required to achieve the desired diameter also decreases. This is because as the diameter of the inhomogeneous phase particles decreases, the number of effective luminescent particles per unit volume increases exponentially for the same doping concentration. Even though a lower doping concentration is used, the number of inhomogeneous phase particles per unit volume increases instead of decreasing. As the luminescent particles act as the second phase, they unavoidably compromise the purity and grain connectivity of the original phase. However, with smaller inhomogeneous phase particle diameters, they can be more uniformly dispersed at the $MgB_2$ particles, significantly reducing their detrimental effects on the purity of the $MgB_2$ phase. As the inhomogeneous phase particle diameter and doping concentration increase, the residual resistivity of the sample (resistivity at 40 K) and the normal-state resistivity of the samples also increase. This is because the introduction of p-n junction luminescent inhomogeneous phase particles leads to an increase in impurities, defects, stress, and dislocations within the superconductor, thereby reducing the material's "cleanliness." Larger inhomogeneous phase particles have a greater negative impact on the sample's performance. At the same time, the *RRR* value ($\rho(300K)/\rho(40K)$) decreases, further confirming that optimizing the diameter of the inhomogeneous phase particles effectively reduces the disruption of grain connectivity by the impurity phase.

Based on Figure 5b, it is evident that the $T_C$ values of the samples listed exhibit improvements compared to the pure $MgB_2$ sample. Notably, the sample incorporating 2 μm-diameter inhomogeneous phase particles at a doping concentration of 0.9 wt% demonstrates the most significant enhancement, reaching a $T_C$ of 1.2 K. In contrast, the optimal doping concentrations for the samples with inhomogeneous phase particle diameters of 0.5 μm, 1.0 μm, 1.5 μm, 3.0 μm, 4.0 μm, 5.0 μm, and 6.0 μm result in $T_C$ enhancements of merely 0.4 K, 0.6 K, 0.8 K, 1.0 K, 1.0 K, 0.6 K, and 0.2 K, respectively. It is noteworthy that as the inhomogeneous phase particle diameter diminishes from 6.0 μm to 0.5 μm, a non-monotonic trend emerges in the $T_C$ performance, peaking when utilizing 2 μm-diameter inhomogeneous phase particles with an optimal doping concentration of 0.9 wt%. Notably, as the inhomogeneous phase particle diameter decreases from 6.0μm to 2.0 μm, the $T_C$ gradually improves despite negligible variations in the doping concentration. This phenomenon can be attributed to the exponential increase in the number of effectively luminescent particles within a unit volume as the inhomogeneous phase particle size diminishes, despite employing lower doping concentrations. Consequently, smaller inhomogeneous phase particles facilitate their uniform dispersion around $MgB_2$ particles, significantly mitigating the purity degradation of the $MgB_2$ phase and preserving crystalline connectivity. Concurrently, as the inhomogeneous phase particle diameter and doping concentration progressively increase, the residual resistivity of the samples at 40 K also rises. This behavior arises from the introduction of luminescent p-n junction inhomogeneous phase, leading to elevated levels of impurities, defects, stresses, and dislocations within the superconducting material, thus compromising its intrinsic "cleanliness." Moreover, larger inhomogeneous phase particle sizes exert a more pronounced negative impact on sample performance. Simultaneously, the RRR value ($\rho(300K)/\rho(40K)$) proportionately diminishes, substantiating the effective attenuation of impurity-induced disruptions to the grain connectivity of the superconducting phase by optimizing the inhomogeneous phase particle size.

Figures 5c and 5d present the experimental results of gradient doping of MgB$_2$ superconductor with 2 μm-diameter inhomogeneous phase particles. The test results demonstrate that lower levels of doping lead to a reduction in the sample's $T_C$, consistent with the majority of doping experiments [50,51]. This phenomenon arises from the introduction of a second phase, which disrupts the material's ordering, resulting in negative gains in terms of connectivity, phonon spectra, and transport properties. At lower concentrations of luminescent inhomogeneous phases, the positive gains from optical field energy are significantly outweighed by the negative effects, leading to a decrease in $T_C$ ($\Delta T_C < 0$). However, when the doping concentration reaches a certain threshold, such as 0.8 wt%, the enhancement effect of the inhomogeneous phase becomes prominent, surpassing the $T_C$ of the pure sample ($\Delta T_C > 0$). At a doping concentration of 0.9 wt%, $\Delta T$ reaches its maximum value of 1.2 K. Further increasing the content of nanomaterial inhomogeneous phases results in a decrease in $\Delta T$ due to the predominance of negative gains from excessive introduction of the second phase, surpassing the positive gains from optical field energy. These characteristics align with previous findings on luminescent inhomogeneous phase doping in oxide materials [27,28,30].

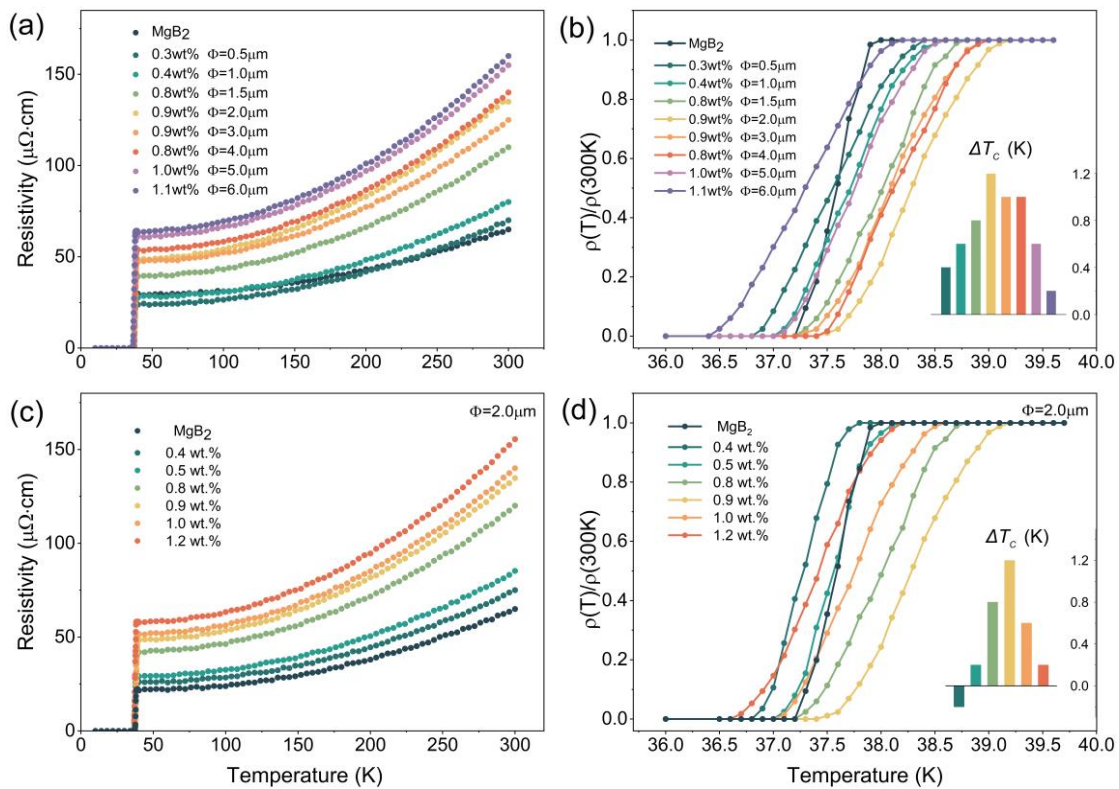

**Figure 5**. Plots of resistivity versus temperature ($T$) for MgB$_2$+ $x$ wt% p-n junction electroluminescent particle samples, (a) normalized resistivity ($\rho$(T)/$\rho$(300K)) versus temperature ($T_C$) for inhomogeneous phase doped samples with different size particles in the temperature range of 10-300 K. (b) Superconducting transitions in the temperature range of 36-40 K for inhomogeneous phase doped samples with different sizes of grains. (c) The plot of normalized resistivity ($\rho$(T)/$\rho$(300K)) versus temperature ($T$) for 2 μm particle size inhomogeneous phase doped MgB$_2$ samples with 0.4 wt%-1.2 wt% concentration

gradient in the temperature range of 10-300 K. (d) Superconducting transitions in the temperature range 36-40 K for 2 μm particle size inhomogeneous phase doped MgB$_2$ samples with 0.4 wt%-1.2 wt% concentration gradient. The inset shows the $\Delta T_C$ of the doped sample compared to the pure MgB$_2$ sample.

Figures 6a and 6b illustrate the influence of different inhomogeneous phase particle sizes on the critical current density ($J_C$) performance of the samples, while Figures 6c and 6d depict the effect of doping concentration variation on the critical current density performance. From Figure 6a, it can be observed that both pure MgB$_2$ and doped samples exhibit a decrease in $J_C$ as temperature increases, consistent with the findings reported in literature [52-54]. The $J_C$ of pure MgB$_2$ at 20 K is measured to be $8.5 \times 10^4$ A/cm$^2$, in agreement with previous studies [55,56]. Using 2 μm-diameter inhomogeneous phase particles with a doping concentration of 0.9 wt%, the sample achieves the optimal critical current performance with $J_C$ of $9.3 \times 10^4$ A/cm$^2$. At lower temperatures, $J_C$ exhibits a gradual decline, while the rate of decline accelerates with increasing temperature. The doping of luminescent inhomogeneous phases enhances the $J_C$ of the samples, with a $J_C$ of 0.9wt% doping sample being 52.8% higher than that of pure MgB$_2$ at $T$ = 36 K. Additionally, when the inhomogeneous phase concentration is 0.4 wt%, the $J_C$ of the sample decreases more rapidly than that of pure MgB$_2$, as shown in Figures 6c and 6d.

In addition, we investigated the impact of inhomogeneous phase particle size on the critical current density ($J_C$) performance of the doped samples, as depicted in Figures 6a and 6b. As the inhomogeneous phase particle size increases from 2.0 μm to 6.0 μm, the critical current performance of the samples gradually decreases. The main mechanism behind the enhancement of MgB$_2$ flux pinning capability is attributed to the precipitation of the introduced p-n junction nanoscale luminescent particles as the second phase around MgB$_2$ particles, forming strong flux pinning centers [37,57-60]. An increase in inhomogeneous phase particle size results in a decrease in the number of luminescent particles per unit volume, and thus a reduction in the number of effective pinning centers, even at the same doping level. Additionally, larger inhomogeneous phase particle sizes can compromise important parameters such as the grain connectivity of the superconducting phase. When the inhomogeneous phase particle size decreases from 2.0 μm to 0.5 μm, the optimal doping concentration also decreases. Although the number of pinning centers per unit volume does not change significantly, the smaller inhomogeneous phase particle size helps reduce the detrimental effects of introducing the second phase on the original superconducting grain structure. However, the critical current performance of the samples also decreases. This observation may seem contrary to the expected trend, but it can be explained by the inherent structural size limitations of the p-n junction luminescent particles. The intrinsic multilayered design restricts the minimum thickness of the luminescent particles to only 1.7 μm, with directional characteristics. Smaller particle sizes contribute to improving various properties of the superconducting material, but excessively small sizes can lead to the destruction of the original luminescent structure, resulting in decreased efficiency or even rendering the second phase non-functional impurities.

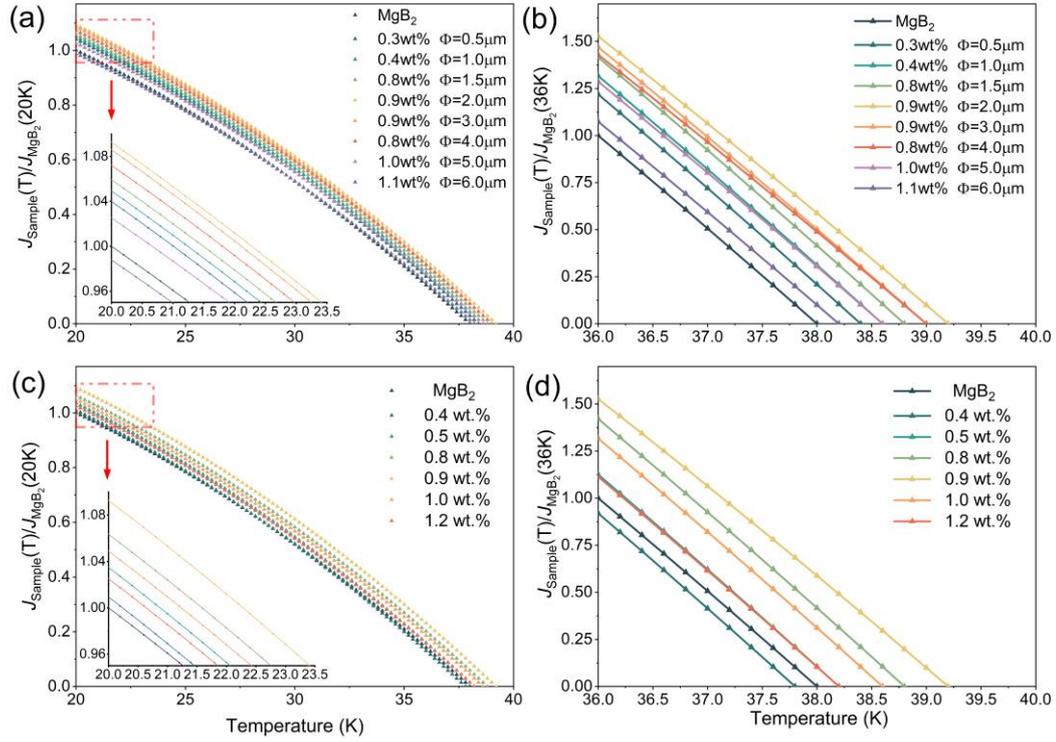

**Figure 6.** Plots of critical current density ($J_C$) versus temperature (T) for MgB$_2$ + $x$ wt% p-n junction electroluminescent particles samples, (a) Normalized critical current density ($J_{Sample}/J_{MgB_2}(20K)$) versus temperature ($T$) for inhomogeneous phase doped samples with different size particles in the temperature range of 20-40 K. (b) Critical current densities in the temperature range of 36-40 K for inhomogeneous phase doped samples with different sizes of grains. (c) Plot of normalized critical current density ($J_{Sample}/J_{MgB_2}(20K)$) versus temperature ($T$) for 2μm particle size inhomogeneous phase doped MgB$_2$ samples with 0.4 wt%-1.2 wt% concentration gradient in the temperature range of 20-40 K. (d) Critical current density of 2μm particle size inhomogeneous phase doped MgB$_2$ samples with 0.4 wt%-1.2 wt% concentration gradient in the temperature range of 36-40 K.

Figure 7 presents the DC magnetization data of pure MgB$_2$ and MgB$_2$ doped with inhomogeneous phase particles. The vertical axis represents the magnetization variation of the materials, and the DC magnetization data allows us to observe the Meissner effect in all samples. As the temperature increases, the Meissner effect weakens and eventually disappears, which is consistent with the findings in literature [34,61-63]. The Meissner effect in pure MgB$_2$ sample vanishes at 37.2 K, while the Meissner effect in the 0.4 wt% inhomogeneous phase MgB$_2$ sample disappears at 36.6 K. For the 0.9 wt% 2.0 μm, 1.0 wt% 3.0 μm, and 0.8 wt% 4.0 μm MgB$_2$ samples, the Meissner effect disappears at temperatures higher than 38.4 K, 38 K, and 37.5 K, respectively. It can be observed that compared to pure MgB$_2$, the diamagnetic performance of the highly doped inhomogeneous phase samples has been significantly improved.

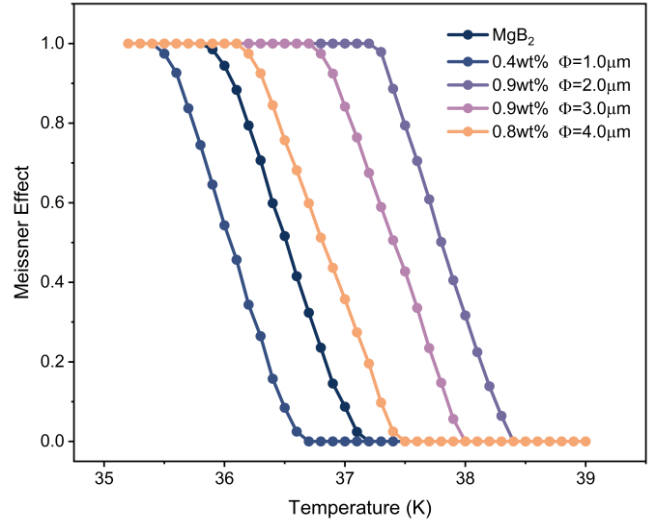

**Figure 7**. DC magnetization data of pure MgB$_2$ and MgB$_2$ doped with inhomogeneous phase

The Raman spectra of MgB$_2$ superconductors doped with different concentrations of p-n junction luminescent inhomogeneous phases, measured at room temperature, are shown in Figure 8. The black data points represent the actual measured data, while the red curve represents the Gaussian fit results [64]. In the Raman spectrum of pure MgB$_2$ sample, the E$_{2g}$ phonon mode appears as the main peak, with a lower frequency of approximately 582.2 cm$^{-1}$, as shown in Figure 8(a). This is consistent with the observations of other experimental groups [65-67]. The E$_{2g}$ phonon mode of pure MgB$_2$ sample exhibits a large linewidth of approximately 200 cm$^{-1}$, which is attributed to strong electron-phonon coupling and phonon-phonon interactions [65,68-70]. For the samples doped with inhomogeneous phases, slight differences can be observed in their Raman spectra compared to pure MgB$_2$ sample. This is mainly due to the presence of the second phase in the samples. The introduction of the p-n junction luminescent inhomogeneous phase induces strong stress in the MgB$_2$ lattice, resulting in disorder and changes in grain connectivity in the sample. After doping with 0.5 wt% Y$_2$O$_3$, the ω and γ of the E$_{2g}$ phonon mode are 583.7 cm$^{-1}$ and 175.4 cm$^{-1}$, as shown in Figure 8(b). The results indicate a hardening of the E$_{2g}$ phonon mode due to Y$_2$O$_3$ doping, which weakens the electron-phonon coupling in MgB$_2$, leading to a decrease in $T_C$ to 37.6 K. The softening of the E$_{2g}$ phonon mode suggests an enhancement of the electron-phonon coupling in the sample, which is beneficial for improving $T_C$ [70-73]. For MgB$_2$ doped with 0.4 wt% inhomogeneous phase with a particle size of 1 μm, 0.8 wt% with a particle size of 2 μm, 0.9 wt% with a particle size of 3 μm, and 0.9 wt% with a particle size of 2 μm, the ω (γ) values of the E$_{2g}$ mode are 564.4 cm$^{-1}$ (226.5 cm$^{-1}$), 557.2 cm$^{-1}$ (237.7 cm$^{-1}$), 552.5 cm$^{-1}$ (242.3 cm$^{-1}$), and 550.6 cm$^{-1}$ (247.1 cm$^{-1}$), respectively. The corresponding $T_C$ values are 38.6 K, 38.8 K, 39.0 K, and 39.2 K, indicating an increase in $T_C$ with the enhancement of the softening effect. These results suggest that the softening of the E$_{2g}$ phonon mode in MgB$_2$ doped with GaN may be the main reason for the enhancement of $T_C$.

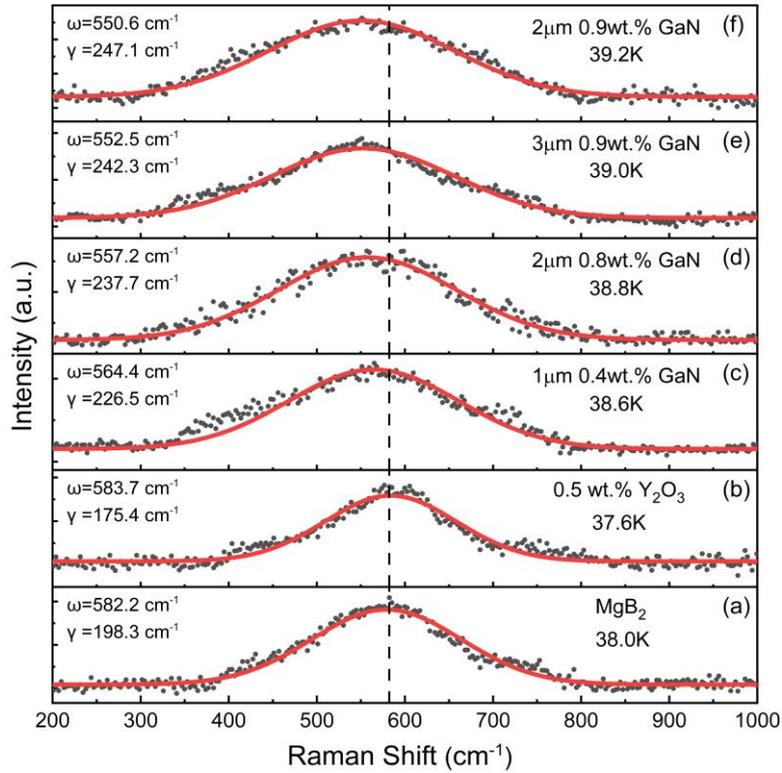

**Figure 8.** Raman spectra of (a) pure $MgB_2$ and $MgB_2$ doped with (b) 0.5 wt% $Y_2O_3$, (c) 1 μm 0.4 wt% GaN, (d) 2 μm 0.8 wt% GaN, (e) 3 μm 0.9 wt% GaN and (f) 2 μm 0.9 wt% GaN

In the aforementioned experiment, we fabricated smart meta-superconductors (SMSCs) composed of conventional $MgB_2$ superconducting matrix and electroluminescent inhomogeneous phase particles as building blocks. The experiment investigated the effects of the size and doping level of the p-n junction electroluminescent inhomogeneous phase particles on the structure and superconducting properties of the SMSCs. The following issues need further elaboration:

**(1) p-n junction electroluminescent inhomogeneous phase particles and chemical doping**

The behavior of modifying the properties of superconducting materials through chemical doping is a widely adopted method [76-79]. For instance, by introducing Ga as a dopant in $MgB_2$, the $T_C$ performance of the sample continuously improves under 400°C heating conditions. As the doping content increases from 1 wt% to 7 wt%, the performance change caused by doping exhibits a monotonic trend, and the $T_C$ of the sample stabilizes at 38.0 K [80]. The green GaN epitaxial wafer used in our experiment is prepared through metalorganic chemical vapor deposition on a sapphire substrate, grown in a vacuum environment above 1300°C. Heating above 1000°C in nitrogen or argon atmosphere may result in very slow decomposition. The sample preparation process in our experiment involves heating at 850°C for 10 minutes, followed by holding at 650°C for 1 hour and slow cooling, which does not meet the decomposition conditions for GaN epitaxial wafers. Therefore, GaN decomposition is not expected to occur in this case. Furthermore, to ensure that the p-n junction particles do not decompose or lose luminescence activity during high-temperature sintering, we also conducted high-temperature sintering experiments on GaN luminescent inhomogeneous phase particles at 850°C for 2 hours. The results indicate that the high-temperature sintering process of the prepared sample does not disrupt the original structure and luminescence performance of the p-n junction particles.

The XRD characterization of the samples indicates that, besides the inherent $MgB_2$ phase, MgO phase, and GaN luminescent phase, there are no other impurity phases present in the samples. This observation verifies that there was no occurrence of GaN decomposition leading to the formation of new compounds or phases during the experimental process. The lattice parameters of pure $MgB_2$ were observed to be a = 3.08532 Å and c = 3.52317 Å, which show good agreement with the standard $MgB_2$ lattice data. When we conducted doping experiments using inhomogeneous phase particles with different contents and sizes, the lattice parameters and diffraction peak intensities of the samples did not show significant shifts or changes, indicating the absence of element substitution (as shown in Table 1). These experimental results also demonstrate that our inhomogeneous phase doping is situated between $MgB_2$ particles.

Elemental doping leads to changes in superconducting performance with a monotonic trend. With increasing doping content, the $T_C$ performance of the samples continuously improves. When the doping content is increased from 1 wt% to 7 wt%, the $T_C$ of the samples reaches 38.0 K, representing a 0.5 K enhancement compared to the pure $MgB_2$ ($T_C$ of 37.5 K) [80]. In our experiment, as the content of the luminescent inhomogeneous phase increases, the various properties of our samples initially improve and then decrease, following the expected trend of our designed SMSCs model. This observation also indicates significant differences between our experimental approach and the principles demonstrated in reference [80].

Furthermore, Raman spectroscopy analysis of the samples indicates that the peak position of the $E_{2g}$ phonon mode shifts with the concentration of the inhomogeneous phase in the samples. As the doping concentration increases, the superconducting critical transition temperature ($T_C$) increases, and the peak of the $E_{2g}$ phonon mode shifts towards lower frequencies. When the doping content exceeds the optimal value, the $T_C$ of the samples decreases, and the peak shifts towards higher frequencies, even surpassing that of pure $MgB_2$ samples. This observation not only demonstrates the performance variation with structural changes but also corroborates the accuracy of the critical transition temperature measurements.

All these findings strongly support the presence of GaN in the form of an inhomogeneous phase in our experiment, ruling out any chemical decomposition. The improvement in superconducting performance can be attributed to the effects of the inhomogeneous phase. This phenomenon is consistent with the results obtained from our previous experiment with red p-n junction AlGaInP and also bears similarities to the effects of $Y_2O_3$ rare-earth doping.

**(2) The universality of inhomogeneous phase in improving critical transition temperature ($T_C$)**

In previous studies, we used $Y_2O_3$ as the inhomogeneous phase to improve the critical transition temperature of $MgB_2$, achieving $ΔT$ of 0.4 K and a transition temperature of 38.6 K [32]. Doping with AlGaInP red p-n junction luminescent inhomogeneous phase resulted in $ΔT$ of 0.8 K and an elevated transition temperature of 39.0 K [35]. Now, with GaN green p-n junction luminescent inhomogeneous phase, $ΔT$ reaches

1.2 K, and the transition temperature increases from 38 K to 39.2 K. It can be observed that the effect of these electroluminescent inhomogeneous phases is not significantly influenced by the material composition but rather depends on the performance of luminescence under an electric field. The enhancement of superconducting performance achieved through the doping of inhomogeneous phases using electroluminescence is a widespread phenomenon.

**(3) Effect of inhomogeneous phase geometry and content on the improvement of transition temperature**

The experimental results mentioned above show that the content of the inhomogeneous phase has a significant impact on the transition temperature. When the content is below a certain value, the enhancement effect is not observed, and the transition temperature decreases. Once a certain concentration is reached, the transition temperature increases compared to the pure sample, and then it reaches a maximum value and decreases as the concentration continues to increase. The entire process of p-n particles is similar to that of $Y_2O_3$ particles, but the luminescence intensity increases, enhancing this effect. Additionally, the geometric size of the inhomogeneous phase exhibits similar behavior, which may be attributed to the luminescence behavior induced by the electric field. Therefore, further research is needed to improve the geometric size and luminescence behavior of the inhomogeneous phase.

**(4) Enhanced effect of green p-n junction luminescent inhomogeneous phase**

From the experimental results, it can be seen that the green p-n junction luminescent inhomogeneous phase not only increases the critical transition temperature $T_C$ and $\Delta T$ but also improves the critical transition current compared to the red particles. The critical current density ($J_C$) increases from 37% of pure $MgB_2$ to 52.8%. Furthermore, the Meissner effect is further enhanced. Therefore, the incorporation of inhomogeneous phases provides a new avenue for improving the performance of intelligent superconductors.

**(5) The physical origin of the increase in critical transition temperature**

Through the construction of electroluminescent inhomogeneous phases doped superconductors, we have achieved the enhancement of superconducting performance. Moreover, we have observed the same enhancement effect using two different luminescent inhomogeneous phases based on $Y_2O_3$ and p-n junctions. However, the origin of this phenomenon has not been fully explained yet. In intelligent superconductors, electroluminescent materials can release energy upon excitation, and the transfer and coupling of this energy have an impact on superconducting performance. However, a theoretical model that fully coincides with experimental phenomena is still lacking. Therefore, based on years of experimental research, we propose the viewpoint that the coupling between evanescent waves generated by electroluminescence and superconducting electrons enhances the critical transition temperature ($T_C$) of superconductors [35]. $MgB_2$, as a conventional BCS superconducting material mediated by phonons, has its superconducting properties closely related to the strength of electron-phonon coupling. By introducing electroluminescent inhomogeneous phases in the form of p-n junctions, the photons generated by the excitation of electroluminescent particles interact with some of the superconducting electrons, resulting in the formation of surface plasmons. These evanescent waves facilitate the unobstructed transmission of superconducting electrons with the same energy, thereby promoting electron interactions in the surface plasmon system. The injection of energy improves the electron pairing state, enhances the superconducting behavior of the material, and raises the critical transition temperature, leading to the formation of the superconducting enhancement effect. In conjunction with the effects brought about by traditional inhomogeneous phase doping, the superconducting properties of doped samples exhibit a dome-shaped evolution with the doping concentration, constituting the smart meta-superconductivity of electroluminescent inhomogeneous phase-doped superconductors. In future experiments, we will endeavor to conduct in-depth research on the ground-state electronic structure, striving to advance the comprehensive understanding of this research field and exploring new approaches to improving the performance of superconducting materials.

## 5. Conclusions

We have constructed smart meta-superconductors (SMSCs) using traditional $MgB_2$ superconducting material as the matrix and p-n junction electroluminescent inhomogeneous particles as the building blocks. In this study, we have systematically investigated the influence of particle size and doping content of the electroluminescent inhomogeneous phase on the structure and superconducting properties of the constructed SMSCs. The results demonstrate a significant enhancement in the performance of our SMSCs samples compared to pure $MgB_2$ samples. Through the optimization effect of electroluminescent inhomogeneous phases, the critical transition temperature ($T_C$) of the samples has increased by 1.2 K, the critical current density ($J_C$) has improved by 52.8%, and the Meissner effect ($H_C$) has exhibited enhanced diamagnetic behavior. The improvement in superconducting performance resulting from the doping of electroluminescent inhomogeneous phases primarily arises from the coupling between evanescent waves generated by the formation of plasmons through electroluminescence and superconducting electrons. The superconducting enhancement effect brought about by the doping of electroluminescent inhomogeneous phases, along with the conventional impurity effects induced by traditional inhomogeneous phase doping, collectively gives rise to the smart meta-superconducting behavior observed in the doped samples. The comprehensive analysis conducted in this study provides valuable insights and guidance for the optimization design and application of superconducting materials. Future research can further explore superior doping materials within a suitable particle size range and investigate the relationship between their microstructural characteristics and performance, thereby advancing the performance and application scope of superconducting materials.

X.Z.; funding acquisition, X.Z. All authors have read and agreed to the published version of the manuscript.

**Institutional Review Board Statement:** Not applicable.

**Funding:** This research was supported by the National Natural Science Foundation of China for Distinguished Young Scholar under Grant No. 50025207.

**Data Availability Statement:** The data presented in this study are available on reasonable request from the corresponding author.

**Conflicts of Interest:** The authors declare no conflict of interest.